\documentclass[journal=jacsat,manuscript=article, keywords]{achemso}

\usepackage{chemformula} 

\usepackage{amsmath}
\usepackage{amsfonts}
\usepackage{amssymb}
\usepackage{color}
\usepackage{changes}

\author{Philipp D'Astolfo}
\affiliation{Department of Physics, University of  Basel, Klingelbergstrasse 82, 4056 Basel, Switzerland}
\author{Xing Wang}
\affiliation{Department of Chemistry, Biochemistry and Pharmaceutical Sciences, University of Bern, Freiestrasse 3, 3012 Bern, Switzerland}
\author{Xunshan Liu}
\affiliation{Department of Chemistry, Biochemistry and Pharmaceutical Sciences, University of Bern, Freiestrasse 3, 3012 Bern, Switzerland}
\altaffiliation{Present address: Key Laboratory of Surface \& Interface Science of Polymer Materials of Zhejiang Province, Department of Chemistry, Zhejiang Sci-Tech University, 928 Second Street, Hangzhou, 310018 China}
\author{Marcin Kisiel}
\affiliation{Department of Physics, University of  Basel, Klingelbergstrasse 82, 4056 Basel, Switzerland}
\author{Carl Drechsel}
\affiliation{Department of Physics, University of  Basel, Klingelbergstrasse 82, 4056 Basel, Switzerland}
\author{Alexis Baratoff}
\affiliation{Department of Physics, University of  Basel, Klingelbergstrasse 82, 4056 Basel, Switzerland}
\author{Ulrich Aschauer}
\affiliation{Department of Chemistry, Biochemistry and Pharmaceutical Sciences, University of Bern, Freiestrasse 3, 3012 Bern, Switzerland}
\author{Silvio Decurtins}
\affiliation{Department of Chemistry, Biochemistry and Pharmaceutical Sciences, University of Bern, Freiestrasse 3, 3012 Bern, Switzerland}
\author{Shi-Xia Liu}
\affiliation{Department of Chemistry, Biochemistry and Pharmaceutical Sciences, University of Bern, Freiestrasse 3, 3012 Bern, Switzerland}
\author{Rémy Pawlak}
\affiliation{Department of Physics, University of  Basel, Klingelbergstrasse 82, 4056 Basel, Switzerland}
\email{remy.pawlak@unibas.ch}
\author{Ernst Meyer}
\affiliation{Department of Physics, University of  Basel, Klingelbergstrasse 82, 4056 Basel, Switzerland}
\email{ernst.meyer@unibas.ch}

\title{Energy dissipation from confined states in nanoporous molecular networks}


\begin{document}

\begin{tocentry}
\centering
\includegraphics[width = 6.5cm]{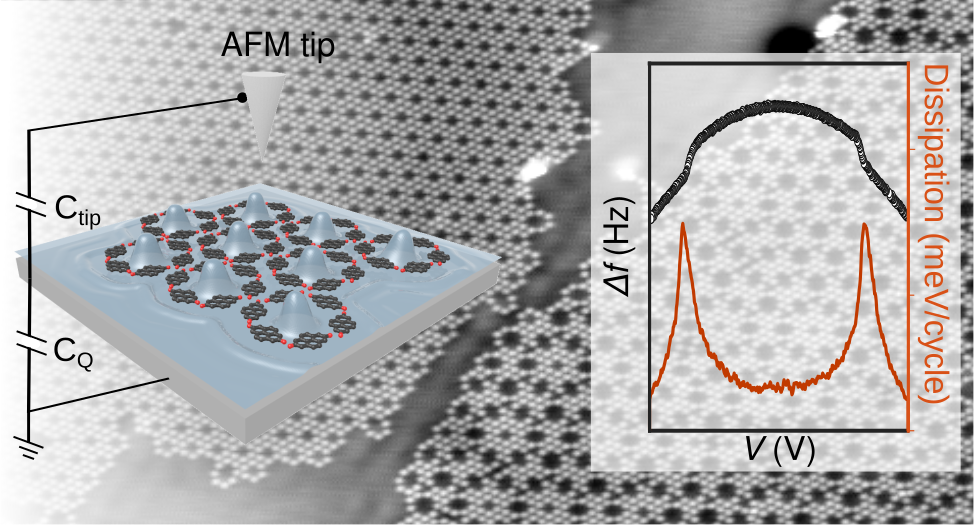}
\end{tocentry}

\begin{abstract}
Crystalline nanoporous molecular networks are assembled on the Ag(111) surface, where the pores confine electrons originating from the surface state of the metal. Depending on the pore sizes and their coupling, an antibonding level is shifted upwards by 0.1 to 0.3 eV as measured by scanning tunneling microscopy. On molecular sites, a down-shifted bonding state is observed, which is occupied under equilibrium conditions. Low-temperature force spectroscopy reveals energy dissipation peaks and jumps of frequency shifts at bias voltages, which are related to the confined states.  The dissipation maps show delocalization on the supra-molecular assembly and a weak distance-dependence of the dissipation peaks. These observations indicate that two-dimensional arrays of coupled quantum dots are formed, which are quantitatively characterized by their quantum capacitances and resonant tunneling rates. Our work provides a method for studying the capacitive and dissipative response of quantum materials with nanomechanical oscillators.

\end{abstract}
\section*{Keywords}
supramolecular assembly, artificial atoms, energy dissipation, quantum capacitance, scanning tunneling microscopy, atomic force microscopy, density functional theory
\section*{Introduction}
Since the quantum corral built in 1993 in IBM Almaden,\cite{crommie_confinement_1993} atomic structures capable of confining Shockley surface states (SS) of noble metals into discrete energy levels (also referred to as artificial atom) are achieved by atom-by-atom manipulation with a scanning tunneling microscope (STM)\cite{Khajetoorians2019,Stilp2021} or by spontaneous assembly of molecules into nanoporous networks.\cite{lobo-checa_band_2009} When a two-dimensional array of artificial atoms is formed, its geometry and symmetry as well as the interaction between the confined states - all depending on the precursor forming the network\cite{Piquero-Zulaica2017} - are key factors in determining the final topology of the band structure. To date, this strategy has been successfully utilized to generate quantum states in artificial lattices such as dispersive electronic bands,\cite{lobo-checa_band_2009,Piquero-Zulaica2017} Dirac fermions\cite{Gomes2012}, fractional behavior in fractal systems,\cite{Kempkes2019} topological edge states\cite{Drost2017} and flat bands.\cite{Slot2017,Gardenier2020,Telychko2021} Interestingly, a partial filling of flat electronic bands can result in a variety of interesting many-body phenomena including Wigner crystallization,\cite{Li2021a} and topological insulating transitions.\cite{Ugajin1994} However, the control of designer quantum states in artificial lattices with a local external electric field has not yet been reported. 
		
Recent progress in dynamic frequency-modulated (FM) atomic force microscopy (AFM)\cite{Giessibl2019} has enabled to obtain spatial resolution at the atomic scale\cite{gross_chemical_2009} and electrostatic force measurements with single-electron sensitivity. The latter has been used to demonstrate single-electron tunneling in individual QDs\cite{Stomp2005,Cockins2010} or the charge-state control of single adatoms\cite{Gross2009} and molecules at surfaces.\cite{Steurer2015,Kocic2015} Experimentally, such charging/discharging phenomena manifest in force spectroscopic measurements versus bias voltage $V$ as drops (peaks) in the cantilever’s resonance frequency shift $\Delta f$ (damping) at constant tip-sample separation $Z$.\cite{Stomp2005} Besides ascertaining tunneling processes in individual QD, $\Delta f(Z)$ spectroscopy can quantify energy loss subsequent to lateral charge transfers between quantized levels\cite{Steurer2015,Scheuerer2020,Berger2020} or tunneling into quantum states in the vicinity of the surface.\cite{Langer2014,Kisiel2018} Recently, Stilp {\it et al.}\cite{Stilp2021} also measured the bonding interaction of an artificial atom made in a quantum corral on Cu(111) with an atomic force microscopy, making AFM an appealing technique for probing the "force" of an artificial atom. Similarly, one could envision to control its charge state or even the filling of a mini-band developed in an artificial superlattice using force versus voltage spectroscopy.\cite{Gross2009}
 
In principle, $\Delta f(Z)$ spectroscopy conducted at low temperature probes tip-sample capacitance including the quantum capacitance $C_{\rm Q}$\cite{Giannazzo2009} arising from a finite and low density of states near the Fermi level $E_{\rm F}$,\cite{Luryi1988} where $C_{\rm Q}$ = $\rho(E)e^2$ relates to the density of states (DOS) $\rho(E) = dn/dE$  with carrier concentration $n$ and $e$ the electronic charge. On noble metals, the Shockley surface state provides a canonical 2D quasi-free electron gas (2DEG) with a quadratic energy dispersion $E(k_{||})$ = $\frac{\hbar^2k_{||}^2}{2m^*}$, where $\hbar$ is the reduced Planck constant,  $k_{||}$ the wave-vector parallel to the surface and $m^*$ the electron's effective mass.  Upon confining surface electrons in two dimensions, the DOS takes the form of a staircase as a result of the formation of dispersive "minibands" in reciprocal space.\cite{Piquero-Zulaica2017,PiqueroZulaica2021} The contribution of the quantum capacitance in AFM measurements of a 2D lattice thus scales as $C_{\rm Q} = \frac{m^* e^2}{\pi \hbar^2}$, that directly encodes the $k$-dispersion relation of the quantized states ({\it i.e.} $m^*$) but also reflects the interaction between artificial atoms.
\begin{figure}[t!]
\centering
	\includegraphics[width = 0.98\textwidth]{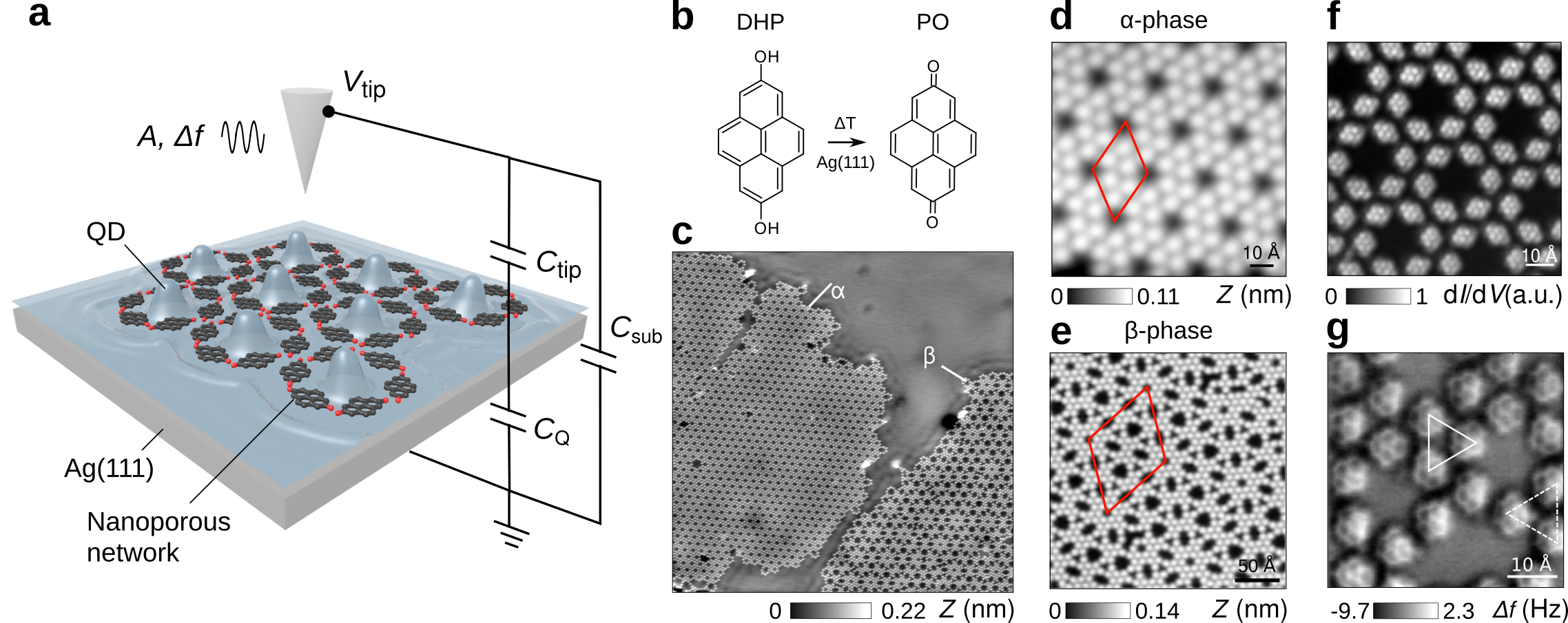}
		\caption{Coupling an AFM to an artificial superlattice.
		{\bf a,} Schematic of the experiment. An oscillating AFM probe  is capacitively coupled to the confined state of surface electrons in a nanoporous molecular network.  $C_{\rm tip}$ and  $C_{\rm sub}$ refers to the capacitance of tip and substrate, respectively. $C_{\rm Q}$ is the quantum capacitance of the "minibands" emerging in the artificial superlattice.
		 {\bf b,} Surface oxidation of 2,7-dihydroxypyrene (DHP) into pyrene-2,7-dione (PO). 
		 {\bf c,} STM overview of the $\alpha$- and $\beta$-assemblies of PO molecules on Ag(111), (scan parameters: $V$ = -0.3 V, $I$ = 1 pA). 
		 {\bf d-e,} STM images of the $\alpha$- and $\beta$-structures, respectively.  (scan parameters: $V$ = -0.13 V, $I$ = 1 pA) 
		 {\bf f,} Bond-resolved STM images of the $\alpha$ network ($A_{\rm mod}$ = 1.5 mV).
			{\bf g,} AFM image with a CO-terminated tip of $\beta$ network displaying hexagonal and octagonal pores ($A_{\rm osc}$ = 50 pm).}
		\label{fig0}
	\end{figure}
\section*{Results/Discussion}
\paragraph{Synthesis of nanoporous molecular networks.}
	Our aim is to employ a low temperature atomic force microscope (AFM) as a local gate capable of probing the capacitance of engineered quantum states in artificial superlattices (Figure~\ref{fig0}a). We created two molecular networks through the thermal evaporation in ultra high vacuum (UHV) of 2,7-dihydroxypyrene (DHP) on the Ag(111) surface. These precursors are oxidized to pyrene-2,7-dione (PO, Figure~\ref{fig0}b) species under experimental conditions as verified by DFT calculations (Supporting Figure 2b). PO molecules spontaneously form nanoporous self-assemblies as a function of the substrate temperature (see Materials and Methods). The topographic STM image of Figure~\ref{fig0}c shows the molecular assemblies, denoted $\alpha$- and $\beta$, where detailed insights are derived from the STM images of Figures~\ref{fig0}d-e and Supporting Figure 1. The $\alpha$-network consists of a hexagonal arrangement with lattice parameter $a_{\alpha} \approx$ 2.9 nm (Figure~\ref{fig0}d). The assembly is governed by cooperative intermolecular hydrogen bonds (Supporting Figure 2a) leading to identical hexagonal cavities of 0.95 nm~diameter and separated from the surrounding ones by two precursors ($\approx$ 1.5 nm). The $\beta$-assembly is a hexagonal lattice of parameter $a_{\beta} \approx$ 6.7-8.7 nm which is composed of trigonal, hexagonal, octagonal and nonagonal cavities. While both networks coexist at the surface, they do not intermix (Figure~\ref{fig0}c).

An atomic understanding of the $\alpha$ and $\beta$ structures is obtained $via$ constant-height AFM and bond-resolved STM imaging\cite{gross_chemical_2009} with CO-terminated tips (Figures~\ref{fig0}f and g) combined with calculations based on a deep-learning neural network (DPNN) potential model (see Materials and Methods and Supporting Text 1). Our calculations (Supporting Figure 2b) reveal DHP molecules to be oxidized to pyrene-2,7-dione (PO) species under experimental conditions. While molecules lie flat in both structures, neighboring PO molecules of the $\alpha$ network interact through C-H$\cdot\cdot$O-C hydrogen bonds between carbonyl groups and peripheral ortho-hydrogens, leading to chiral trimers of PO molecules (plain triangle, Figure~\ref{fig0}g and Supporting Figure 2a). A similar bonding motif is observed for hexagonal pores in the $\beta$-assembly (Figure~\ref{fig0}g), whereas trigonal, octogonal and nonagonal cavities result from different H-bonding motifs between PO molecules. As a result, the $\beta$-structure contains four cavity geometries as reproduced by our DPNN-based calculations (Supporting Figure 2c).
\begin{figure}
\centering
\includegraphics[width = 0.95\textwidth]{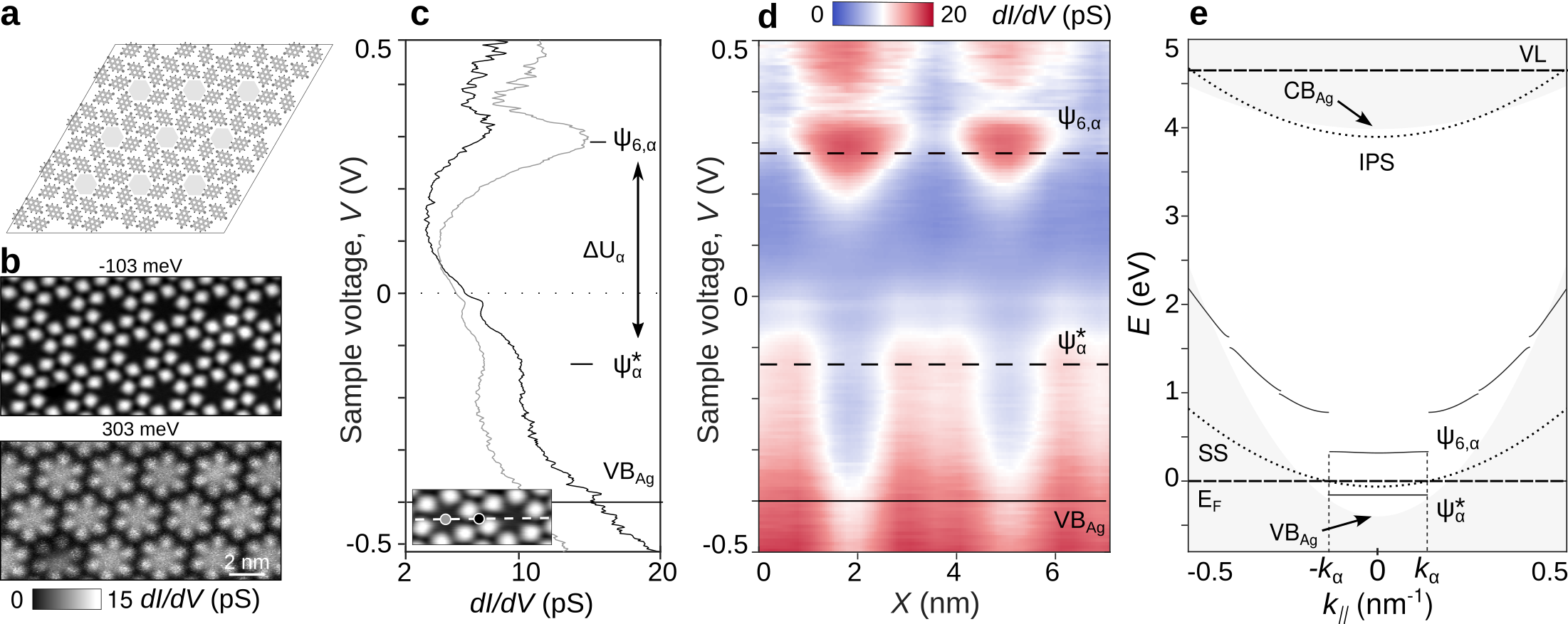}
\caption{Electronic structure of the $\alpha$ superlattice.
{\bf a,} Structure of the $\alpha$ lattice on Ag(111) obtained by DNPP calculations and  
{\bf b,} LDOS(X,Y) maps at $\psi^*_{\alpha}$ = -103~mV and $\psi_{6,1}$ = 303~mV showing the spatial localization of the molecule-induced bound state and confined states in the superlattice, ($A_{\rm mod}$ = 6 mV).
{\bf c,} d$I$/d$V$(V) spectra acquired at the center of a pore (gray) and above a molecule (black) (see inset, scan parameters: $V$ = -0.15 V, $I$ = 1 pA). Resonance peaks at 302 mV and -102 mV corresponds to the $\psi_{6}$ and $\psi^*_{\alpha}$ states  separated by a gap $\Delta U$ = 430 mV.  
{\bf d,} d$I$/d$V$(X,V) cross-section acquired along two $\alpha$-pores as marked by the dashed line in the inset. 
{\bf e,} Schematic of the projected electronic band structure of the $\alpha$ QD superlattice. The $E$(k) dispersion considering the $\alpha$- lattice in the form of a Kronig-Penney model is plotted in gray.}
\label{fig2}
\end{figure}
\paragraph{Electronic structure of the superlattices.}	
We next employed scanning tunneling spectroscopy (STS) to probe the local density of states (LDOS) and wave-functions of both superlattices (Figures~\ref{fig2} and 3). On Ag(111), the unperturbed surface state appears at -67 meV (Supporting Figure 1g) with a quadratic energy dispersion $m^*$ = 0.397$\times m_0$;  the wave vector $k$ is parallel to the surface with a Fermi wave vector of $k_{\rm F}$ = 2$\pi/\lambda_{\rm F}$ $\approx$ 0.8 nm$^{-1}$ and $\lambda_{\rm F}$ $\approx$ 7.8 nm. The top of the valence band is located at $E_0$ = -400~meV. Figure~\ref{fig2}c shows d$I$/d$V$ spectra acquired at the center of an $\alpha$-pore and above a PO molecule of the assembly (see inset). A resonance peak is observed at $\psi_{6,1}$ = +300~meV corresponding to the first eigen-state $\psi_{i = 6, n = 1}$ ($i$ is the edge number of the polygonal cavity, $n$ is the quantization integer) due to quantum confinement of surface electrons in the cavity. The resonance peak near 0 V is attributed to a signature of the electronic state of the tip atom.\cite{Kwapinski2010}  Below $E_{\rm F}$, a resonance state at -130 meV denoted $\psi^*_{\alpha}$ is observed at molecule sites (black spectra in Figure~\ref{fig2}c and d$I$/d$V$ map of Figure~\ref{fig2}b), which is attributed to the formation of a molecule-induced bound state lowering the surface state energy.\cite{Limot2005} The $\psi_{\alpha}$ and $\psi^*_{\alpha}$ resonance peak persists at the molecule and pore sites (gray and black spectra in Figure~\ref{fig2}c), respectively, which indicates the concomitant formation of filled ($\psi_{\rm \alpha}^*$) and unoccupied ($\psi_{\rm \alpha}$) minibands in the superlattice upon the quantum confinement of surface electrons in cavities. We also conclude that the strong hybridization between artificial atoms of the superlattice leads to the splitting of the confined states into a bonding ($\psi_{\rm \alpha}^*$) and antibonding states ($\psi_{\rm \alpha}$) similar to Refs.\cite{Sperl2009,Seufert2013,Peng2021}  A gap of $\Delta U$ = 430 meV is thus deduced throughout the $\alpha$-lattice. The d$I$/d$V$~mapping at the $\psi_{6}$ energy reveals the spatial localization of the $\psi_{6}$ wave-function (Figure~\ref{fig0}c) as bright contrasts at the center of the cavities. 
Figure~\ref{fig2}d shows a LDOS(X,V) cross-section taken along two neighboring cavities (see dashed line of inset of Figure~\ref{fig2}c) demonstrating the high fidelity of both $\psi_{6,\alpha}$ and $\psi^*_{\alpha}$ states throughout the porous array.  

Figure~\ref{fig2}d depicts the projected band structure of the $\alpha$-array considering the $\psi_{\rm 6}$ single-particle levels as compared to the top of the valence band (VB$_{\rm Ag}$  marked by an arrow at -0.4 eV), the bottom of the conduction band (CB$_{\rm Ag}$ marked by an arrow at +4.0 eV),\cite{Cui2014} the surface state (SS at -67 meV) and the image potential state (IPS at 3.9 eV) of Ag(111). The latter is located at +3.9 eV as confirmed by measuring field-emission resonance states by tunneling d$Z$/d$V$ spectroscopy (Supporting Figure 9 and Supporting Text 3). We reproduce the $E$(k) dispersion relation of the eigen-states by taking into account the dimensions of the $\alpha$ lattice in an one-dimensional array of finite quantum wells in the form of a Kronig-Penney potential $U$(X), whose value is 0.8 eV at the top of the wall and the SS energy for the well bottom. As shown in Figure~\ref{fig2}e, $\psi_{6}$ emerges at +0.30 eV with a gap opening at the Brillouin zone boundaries $k_{\alpha}$. The $\psi_{\rm 6}$ extends in $k$-space between $\pm k_{\alpha}$ = $\pi$/$a_{\alpha}$ = 1.09 nm$^{-1}$ states  with a dispersion arbitrarily approximated to $m^*$ = 0.45 in relative agreement with ARPES measurements on similar porous systems on silver.~\cite{Piquero-Zulaica2017}%

\begin{figure}[t!]
\centering
\includegraphics[width = 0.95\textwidth]{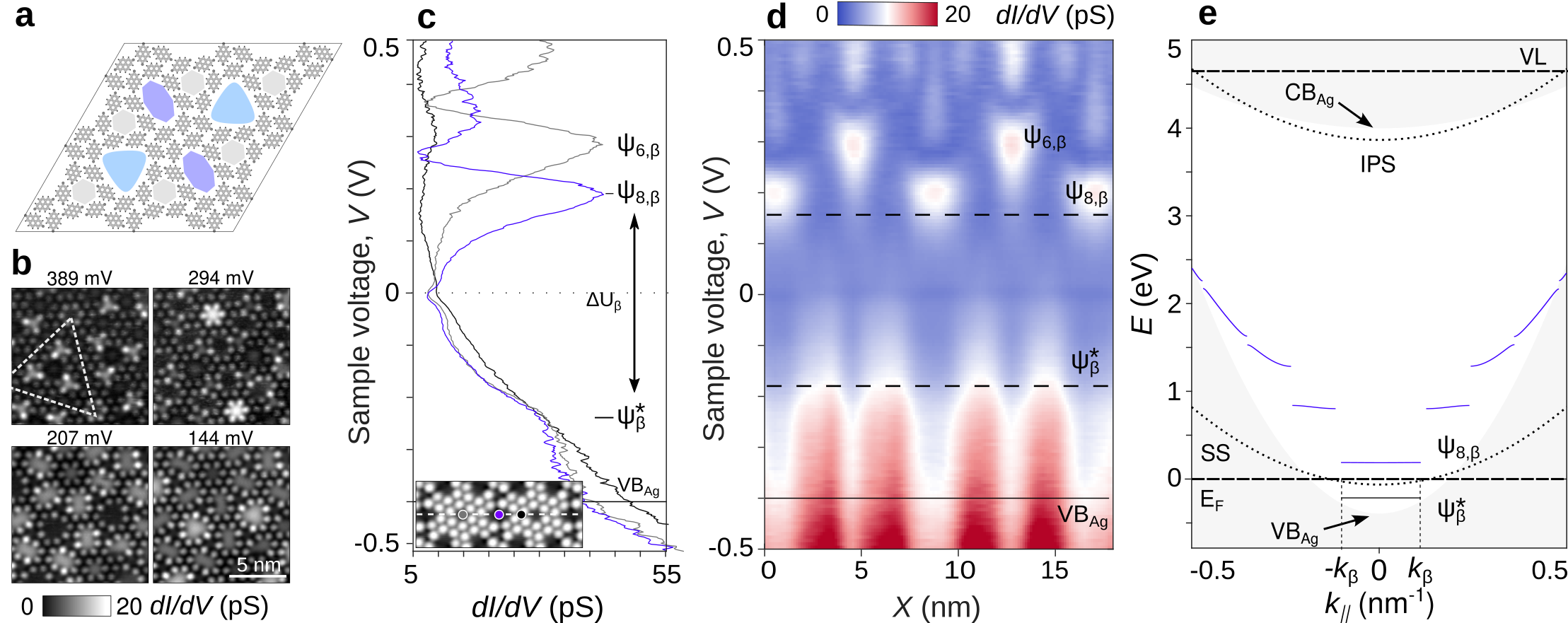}
\caption{Electronic structure of the $\beta$ superlattice.
{\bf a,} Structure of the $\beta$-superlattice on Ag(111) obtained by DNPP calculations containing trigonal, hexagonal and octogal cavities colored in pale blue,  gray and dark blue,  respectively. 
{\bf b,} Series of LDOS(X,Y) mapping revealing the spacial confinement of the confined states at the corresponding energies ($A_{\rm mod}$ = 6 mV). Resonance states of the 3-, 6-, 8- and 9-member pores are $\psi_3$ = 389~mV, $\psi_6$ = 294~mV, $\psi_8$ = 207~mV and $\psi_9$ = 144~mV, respectively. 
{\bf c,} d$I$/d$V$(V) point-spectra acquired at the center of hexagonal (gray), octogonal (blue) pores and above a molecule (black).
{\bf d,} LDOS(V,X) cross-section taken along hexagonal and octagonal pores (see inset, scan parameters: $V$ = -0.15 V, $I$ = 1 pA).
{\bf e,} Schematic of the projected electronic band structure of the $\beta$-superlattice on Ag(111). The $E$(k) relation considers each pore dimension in the Kronig-Penney model with $m^*$ = 0.45.\label{fig3}}
\end{figure}

We also investigated the confined states induced by the polygonal cavity (i.e. $i$ = 3, 6, 8 and 9) of the $\beta$-superlattice, which are depicted in pale blue, gray and dark blue in the structure obtained by DPNN calculations (Figure~\ref{fig3}a), respectively. The four pore sizes and geometries modulate the confinement landscape leading to four different eigen-energies $\psi$ identified by d$I$/d$V$ spectra acquired in each cavity (Figure~\ref{fig3}c and Supporting Figure 7) at $\psi_{3}$ = 389~meV, $\psi_{6}$ = 295~meV, $\psi_{8}$ = 205~meV and  $\psi_{9}$ = 145~meV, respectively. Figure~\ref{fig3}b shows the wave-function mapping of these eigen-energies $\psi$, again demonstrating a high fidelity of the confined states throughout the nanoporous assembly. This observation is further confirmed by the LDOS(X,V) cross-section (Figures~\ref{fig3}d) acquired along hexagonal and octogonal cavities (see inset). 
 In analogy to the $\alpha$-lattice, a molecule-induced bound state $\psi^*_{\beta}$ emerges at $V$ = -190 mV (Figures~\ref{fig3}c and d), where its shift compared to $\psi^*_{\alpha}$ likely results from different bonding motifs in the $\beta$-phase. A gap $\Delta U$ = 335 meV is estimated between the $\psi^*_{\beta}$ band and the lowest confined eigen-state $\psi_{9}$. The corresponding projected band structure of the $\beta$-superlattice (Figure~\ref{fig3}e) is approximated by lowest $\psi_{\rm 8}$ eigen-values (= +200 meV) and extends in $k$-space according to a 1D Kronig-Penney (KP) model between $\pm k_{\beta}$ = $\pi$/$a_{\beta}$ = 0.5 \AA$^{-1}$ with $m^*$ = 0.45~$m_0$.

\begin{figure}[t!]
		\includegraphics[width=0.90\textwidth]{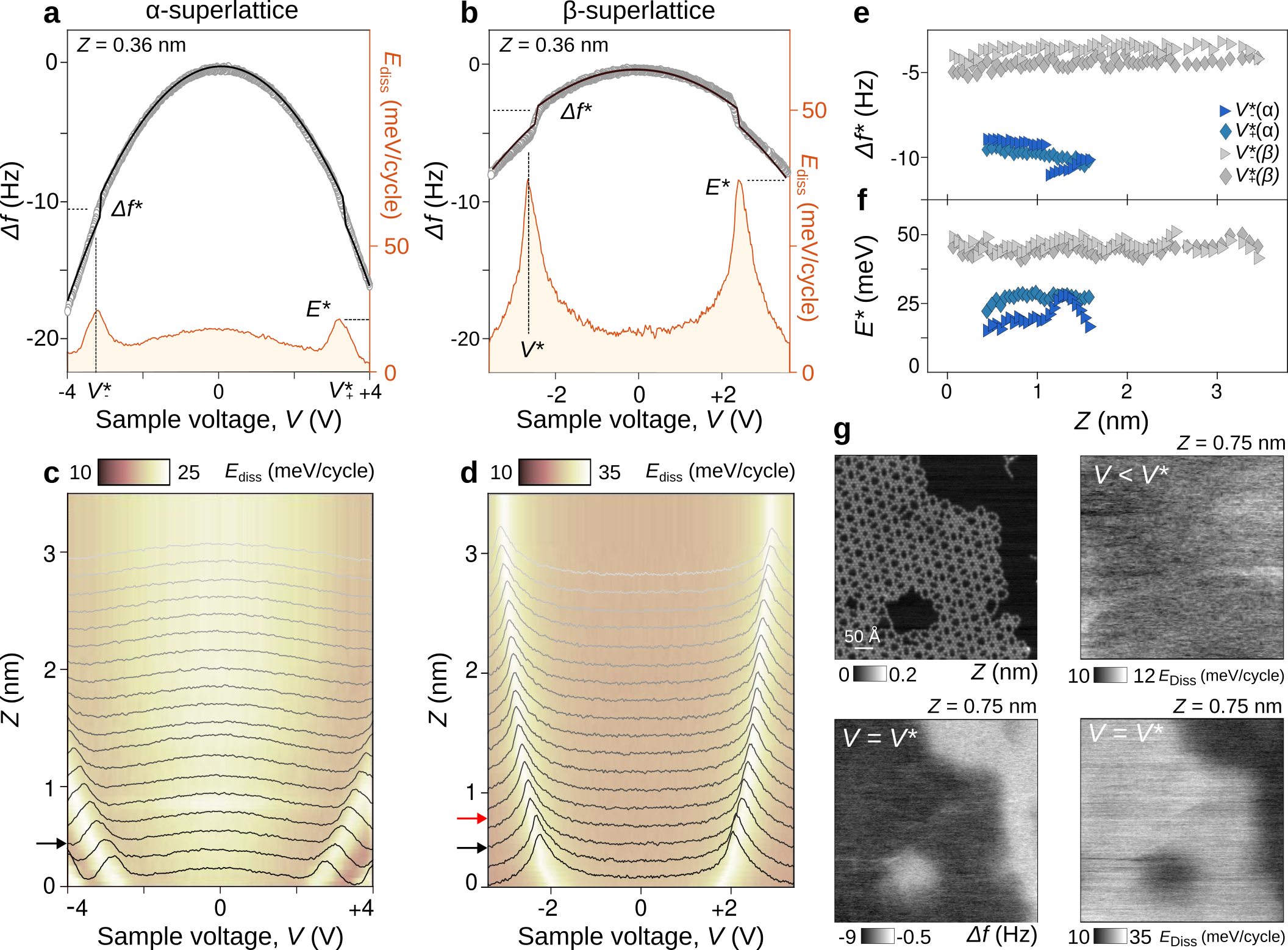}
		\caption{Dissipation spectroscopy above the $\alpha$ and $\beta$ superlattices.
{\bf a-b,} $\Delta f$(V) point-spectra (gray) and associated dissipation $E_{\rm diss}$(V) (orange) spectra of the $\alpha$- and $\beta$-arrays for a relative tip-sample distance of $Z$ = 0.36 nm ($A_{\rm osc}$ = 70 pm). The jump position in the $\Delta f$(V), denoted as $\Delta f^*$, are associated to dissipation peaks $E^*$ at threshold voltage $V^*$.
{\bf c-d} Dissipation map $E_{\rm diss}$(V,Z) acquired above the $\alpha$- and $\beta$-superlattice. The bright lines corresponds to the evolution of $E^*$ peaks as a function of $Z$. Single $E_{\rm diss}$(V) spectrum are superimposed in gray as a guide for the eye. Black arrows show the relative tip-sample distance $Z$ of the spectra shown in {\bf a} and {\bf b}. The red arrow in {\bf d} refers to the tip-sample distance at which dissipation maps of {\bf g} were acquired.
{\bf e-f,} $\Delta f^*$ and $E^*$ magnitudes for increasing $Z$. Blue and gray markers refers to the $\alpha$ and $\beta$ superlattices, respectively. Diamond and triangles corresponds to positive and negative voltage $V$, respectively. 
{\bf g,} Topographic STM image of the $\beta$-superlattice as compared to $\Delta f$(X,Y) and $E_{\rm diss}$(X,Y) maps at $V$ = $V^*$ . The maps show that force/dissipation signals are only detected above the molecular network. The $E_{\rm diss}$(X,Y) map obtained for $V \neq V^*$ at the same tip-sample distance and $Z$ = 0.5 nm show no mechanical dissipation ($A_{\rm osc}$ = 80 pm).}
\label{fig4}
\end{figure}

\paragraph{Quantum capacitance and mechanical dissipation probed by force spectroscopy.}
Next we discuss the detected mechanical dissipation above the artificial lattices using force-voltage spectroscopy with silver-coated tips.\cite{Stomp2005,Cockins2010} Figures~\ref{fig4}a-b show exemplary $\Delta f$(V) (gray) and $E_{\rm diss}$(V) (orange) point-spectra at fixed tip-sample separation $Z$ = 0.36 nm above the $\alpha$- and $\beta$-superlattices, respectively. Both $\Delta f$ and $E_{\rm diss}$ curves show abrupt transitions denoted $\Delta f^*$ and $E^*$ at positive and negative threshold voltages $\pm V^*$, which are absent on pristine Ag(111) sample (Supporting Figure 10). At first glance, the $\Delta f^*$ steps and $E^*$ peaks resemble charging/discharging events as encountered in single QD such as molecules\cite{Steurer2015,Kocic2015,Berger2020} indicating a charge injection/extraction in the minibands of the superlattices similar to Ref.\cite{Kisiel2018} To better rationalize this, we reproduce the $\Delta f(V)$ parabolas (black curves in Figures~\ref{fig4}a and b) using a model that accounts for the quantum capacitance $C_{\rm Q}$ and the tip and substrate capacitance, $C_{\rm tip}$ and $C_{\rm sub}$ (Supporting Text 6). $C_{\rm Q}$(V) is expressed as a Heaviside function which reflects the staircase LDOS of the 2D-superlattice with steps at -$\psi^*$ and +$\psi_{\rm i,1}$ (See Supporting Text 5) as determined by tunneling spectroscopy. The  fit agreement allows to infer $C_{\rm Q}$ equal to 2.2 aF/nm$^2$ and 18.3 aF/nm$^2$ for the $\alpha$ and $\beta$ QD-superlattice, respectively. 

The $\Delta f^*$ jumps are accompanied by an increase of $E^*$ up to 20 meV/cycles and 40 meV/cycles for the $\alpha$- and the $\beta$-assembly, as compared to the unperturbed oscillator (15 meV/cycles) recorded at the compensated contact potential difference (CPD). This minute amount of dissipation detected by the small oscillation amplitude of the actuator~\cite{Kawai2012} suggests electron charging/discharging events from the superlattices,\cite{Cockins2010,Kisiel2018} which is absent on pristine Ag(111) (Supporting Figure 10). Note also that the dissipation magnitude is likely overestimated because of a possible apparent dissipation from the non-linear behavior of our instruments, which was not corrected in our experiments as described in Ref.\cite{Labuda2011} The $E_{\rm diss}$(Z,V) cross-section (Figures~\ref{fig4}c-d) shows bright lines associated to the $E^*$ peaks as a function of tip-sample separation $Z$. The $E^*(Z)$ dependency results from the tip ($C_{\rm tip}$) and sample ($C_{\rm sub}$) capacitances as typically observed in force spectroscopic measurements. The values of the threshold voltage $V^*$ depends on the lever arm defined as $\kappa$ = $C_{\rm tip}/(C_{\rm sub} + C_{\rm tip})$ such as $V$ = $\psi/\kappa$ (Supporting Text 4).\cite{Cockins2010} Note also that the maximum voltage $V_{\rm tip}$ applicable to the tip in our experiments is $\pm$ 4 V which corresponds to the energetic position of the image potential state (IPS) identified by field-emission resonance tunneling spectra (Supporting Figure 9) that is known to induce mechanical dissipation by charge injection between tip and sample thus possessing strong distance dependence.\cite{Yildiz2019}  

In Figures~\ref{fig4}e-f, $\Delta f^*$ and $E^*$ (triangle vs. square) of $\alpha$- (blue markers) and $\beta$- (gray marker) superlattice are plotted as a function of tip-sample distance $Z$. Their magnitudes are constant for both $V$ polarities as $Z$ increases. This observation is in stark contrast with previous works assessing Coulomb blockade in 0D-systems (such as quantum dots or single molecules),\cite{Cockins2010,Steurer2015,Kocic2015,Scheuerer2020,Berger2020}~ where both $\Delta f^*$ jumps and $E^*$ peaks are dictated by single-electron tunneling between tip and quantized levels. There, the dissipation directly scales with the tunneling rates $\Gamma$ and decays with increasing $Z$\cite{Berger2020} which is in contrast to our observation. This suggests that the tip gating in our system induces charge fluctuations between the quantum states of the artificial lattice ($i.e.$ the bonding or the antibonding states) and the electron reservoir of the substrate but no direct charge transfers between tip and sample. This is followed by a variation of the quantum capacitance of the system detected by our probe and a dissipative response. Another difference is that the DOS in 0D systems adopts a fully discretized delta distribution resulting in the observation in constant-height $\Delta f^*/E^*$ maps of Coulomb rings,\cite{Cockins2010,Kocic2015,Scheuerer2020,Berger2020} whose diameter depends on the local (X,Y,Z) position of the AFM tip with respect to the QD. 

Figure~\ref{fig4}g shows a series of exemplary  $\Delta f^*/E^*$ maps acquired at constant-height $Z$ = 0.75 nm (position which is marked by a red arrow in Figure~\ref{fig4}d) above the $\beta$ superlattice. The dissipation map at $V$ = $V^*$ (bottom right panel) reveals no ring features but instead a strong homogeneous dissipation background above the molecular assembly that vanishes on the Ag(111) surface, implying that the dissipation arises from charging/discharging states induced by the molecular lattice. When $V_{\rm tip} \neq V^*$, no site-dependent dissipation is observed between tip and sample (top right panel), which proves that any dissipation signal detected at such large tip-sample distance is not related to topographic variations in the molecular lattice. The absence of Coulomb rings at $V$ = $V^*$ further confirms the two-dimensional nature of the electron confinement in the $\alpha$- and $\beta$-lattices, giving rise to delocalized wave-functions throughout both superlattices (Supporting Figure 8) contrarily to an array of weakly coupled quantum dots. This is in strong analogy to our previous work reporting energy dissipation above a reduced reconstruction of SrTiO$_3$ where the observed dissipation peaks were attributed to tip-induced charge and spin state transitions in an ensemble of quantum-dot-like entities formed by oxygen vacancies.\cite{Kisiel2018}

From the dissipation data, we also inferred the charge transfers between the superlattice's eigen-states and the substrate, resulting in the filling (unfilling) of the $\psi$ ($\psi^*$) minibands. As charges may switch back and forth at the voltage threshold $V^*$, we estimated the tunneling rate $\Gamma$ to be in the order of about 65 kHz and 420 kHz for the $\alpha$- and $\beta$-superlattices, respectively (see Methods, Supporting Text 5 and Figure 14). Note also that the overestimation of the dissipation magnitude in our experiments might lead to a substantial decrease of these extracted tunneling rates (Eq. 17 in Supporting Text 5). We assume that these values might refer to resonant tunneling between artificial atoms or between the minibands of the superlattice.  Since electrons are more confined in the $\alpha$-lattice, the frequency of fluctuating currents is expected to decrease as compared to the $\beta$ one.
 
\section*{Conclusions}
In summary, we characterized using low temperature (4.8 K) scanning tunneling microscopy (STM) and atomic force microscopy (AFM) combined with density functional theory (DFT) two crystalline nanoporous networks formed on a Ag(111) surface by self-assembling pyrene-2,7-dione molecules. Scanning tunneling spectroscopy shows the confinement of the surface electrons into the pores of the molecular networks, leading to levels shifted upwards by 0.1 to 0.3 eV above the Fermi level. As a result of the strong coupling between these artificial atoms, the confined levels form bonding/antibonding states, leading to the formation of an occupied and unoccupied band delocalized over the lattice.\cite{Seufert2013,Peng2021} Force (dissipation) versus voltage spectroscopy acquired above the lattices systematically shows steps (peaks) at both voltage polarities similar to charging/discharging events in 0D quantum dots by local tip gating. We interpret this phenomena as a change of the band occupancy induced by electrostatic gating from the AFM tip, which is rationalized using a capacitance model that includes the capacitance of the substrate and tip as well as the quantum capacitance of the confined states. While STM is known to be very useful for the design and spectral characterization of designer quantum states in artificial lattices,\cite{Khajetoorians2019} our results thus demonstrate that AFM could serve as complementary technique enabling to investigate exotic electronic effects induced by electrostatic gating and quantify their quantum capacitance. We foresee that AFM spectroscopy will not only allow the local gating of artificial lattices on metals but also induce quantum phase transition in dual-gated hetero-structure devices where back-gate voltage can be additionally applied.\cite{Kim2021}

\section*{Methods/Experimental}
\paragraph{Molecule synthesis.}
2,7-Dihydroxypyrene (DHP) was prepared according to literature procedures.\cite{crawford2012synthesis}

\paragraph{Sample preparation.} 
An Ag(111) single crystal purchased from Mateck GmbH was cleaned by several sputtering and annealing cycles in ultra-high vacuum (UHV). DHP molecules were evaporated from a quartz crucible onto the silver substrate annealed at $\approx$ 500~K. The evaporation rate was controlled using a quartz micro-balance.
	 
\paragraph{STM/AFM experiments.} 
STM/AFM experiments were carried out at 4.8 K with an Omicron GmbH low-temperature STM/AFM operated with Nanonis RC5 electronics. We used commercial tuning fork sensors in the qPlus configuration ($f_0$ = 26~kHz, $Q$ = 7000-25000, nominal spring constant $k$ = 1800~N m$^{-1}$). The constant-height AFM images were acquired with CO-terminated tips using the non-contact mode with oscillation amplitude $A_{\rm osc}$ of 50 pm. The AFM spectroscopy has been carried out with oscillation amplitudes of about 70 pm using silver-coated tips, which we obtain by gently indenting the tip apex to the silver substrate prior to experiments.  Differential conductance measurements were carried out with the lock-in technique (lock-in frequency $f$ = 540~Hz, modulation amplitude $A_{\rm mod}$ = 6 mV). 

\paragraph{DFT calculations.}
Density functional theory (DFT) calculations were performed with the Quickstep code \cite{vandevondele_quickstep:_2005} within the CP2K package, using a mixed Gaussian and plane waves basis set, the Goedecker, Teter, and Hutter (GTH) pseudopotentials,\cite{goedecker_separable_1996} and a GGA-PBE\cite{perdew_generalized_1996} exchange-correlation functional including self-consistently the van der Waals (vdW) interaction. We used a plane-wave basis energy cutoff of 500 Ry. To deal with the metallic configurations, we adopted a 0.22 eV Fermi Dirac smearing of the occupation number (electronic temperature 2500 K) around the Fermi energy. The Ag(111) substrate was modeled as a periodically repeated slab of four layers, adding a vacuum gap of ~15 {\AA} between the adsorbed molecule and the bottom layer of the slab above. Structural relaxations were considered completed when atomic forces reached 0.02 eV/{\AA}. Depending on the lateral dimension of the slab cell (a), we used k-point grids of $k\times k \times 1$ to sample the Brillouin-zone and made sure that $k\times a>50$.

The relative stability of the supported molecules compared to their gas phase counterparts is given by (taking DHP as an example):
\begin{equation}
\label{e:relative-stability}
\Delta G_{\mathrm{DHP}/(111)} = G_{\mathrm{DHP}/(111)} - G_{\mathrm{DHP}} - G_{\mathrm{(111)}},
\end{equation}

\noindent where $G_{\mathrm{DHP}/(111)}$, $G_{\mathrm{(111)}}$ and $G_{\mathrm{DHP}}$ are the free energies of the surface with an adsorbed DHP molecule, a clean Ag (111) surface and a gas phase DHP molecule, respectively. 

The phase diagram of adsorbed trimers was calculated based on the method of \textit{ab initio} atomistic thermodynamics.\cite{reuter_first-principles_2003} The Gibbs free energy in gas phase at temperature $T$ and partial pressure $P$ is given by:
\begin{equation}
  G(T, P) = E^{\mathrm{DFT}} + E^{\mathrm{ZPE}} + \Delta G(T, P^0) + k_{\mathrm{B}}T\mathrm{ln}(P/P^0),
  \label{e:muo2-3}
\end{equation}

\noindent where $E^{\mathrm{DFT}}$ is the energy calculated by DFT at 0 K, $E^{\mathrm{ZPE}}$ is the zero point energy, $P^0$ is the standard pressure, and $\Delta G(T, P^0)$ includes the contributions from translational, rotational, vibrational and electronic free energy terms of the species under consideration. The detailed derivation for $\Delta G(T, P^0)$ can be found elsewhere.\cite{reuterabinitioTD} These were implemented in the Atomic Simulation Environment (ASE) Python package.\cite{ase-paper} The change of the Gibbs free energy of the solid phase with $T$ and $P$ is much smaller compared to the gas phase, and it is therefore neglected in this study.

\paragraph{Deep learning NN potential model.}
In the present work, the deep learning neural network (DPNN) potential was constructed using the DeepPot-SE model proposed by Zhang et al.\cite{zhang2018deep} In this model, 
the total potential energy ($E$) of a system is the sum of atomic energies ($E_i$) depending on the local environment of atom $i$ within a smooth cutoff radius $R_c$. $E_i$ is constructed in two steps. First, for each atom a set of symmetry-preserving descriptors is constructed. Next, this information is given as input to a DNN, which returns $E_i$ as the output. The additive form of $E$ naturally preserves the extensive character of the potential energy. The NN potential was trained with the DeepMD-kit package.\cite{zhang2018deep} The cutoff radius smoothly decays from 5.5 {\AA} to 6.5 {\AA}. We used three hidden layers with (25, 50, 100) nodes/layer for the embedding network and three hidden layers with (240, 240, 240) nodes/layer for the fitting network. The learning rate decays from $1.0 \times 10^{-3}$ to $3.5 \times 10^{-8}$. The prefactors of the energy and the force terms in the loss function change from 0.02 to 1 and from 1000 to 10, respectively.

\paragraph{Dissipation spectroscopy.}
We simultaneously acquired  the frequency shift $\Delta f(V)$ and the excitation amplitude $A_{\rm exc}$ required to constantly keep  the oscillation amplitude of the tip in interaction with the sample  above  the $\alpha$- and $\beta$-lattices at the tip-sample separation $Z$. The dissipated energy per oscillation cycle $E_{\rm diss}$ is extracted from $A_{\rm exc}$ with the formula:
	\begin{equation}\label{eq:diss}
		E_{\rm diss} \approx \frac{\pi k A^2}{Q} \left[ \frac{A_{\rm exc} - A_{\rm Exc,0}}{A_{\rm exc,0}} \right]
	\end{equation}
where $k$ is the tuning fork stiffness, $A$ the oscillation amplitude of the tip and $Q$ the intrinsic quality factor~\cite{Anczykowski1999}. $A_{\rm exc, 0}$ refers to the excitation amplitude at the contact potential difference far from the surface. The dissipation maps $E_{\rm diss}(Z,V_{\rm tip})$ consist of 58$\times$512 and 72$\times$512 pixels$^2$, respectively. We extract the tunneling rates $\Gamma$ from the $E_{\rm diss}(\Delta f)$ plots  using the formula:\cite{Zhu2008,Cockins2010}
\begin{equation}
\label{eq:zhu}
\Gamma = 4 \pi^2 k A^2 \frac{\delta f}{E_{\rm diss}}
\end{equation}
with $\delta f$ is the full width at half maximum of the $E_{\rm diss}(\Delta f)$ curve and $E_{\rm diss}$ the maximum dissipation (Supporting Text 5 and Figure 14).

\bibliography{references}%

\section*{Author contributions}
R.P., S.D., S.-X.L. and E.M. conceived the experiments. X.L., S.-X.L. and E.M. synthesized the monomer. P.D. and R.P. performed the STM/AFM measurements. U.A. and X.W. performed the DFT calculations. R.P., M.K., A.B. and E.M. analyzed the data. R.P. wrote the manuscript. All authors discussed on the results and revised the manuscript.

\section*{Competing interests}
The authors declare no competing financial interests. 

\begin{acknowledgement}
We thank the Swiss National Science Foundation (SNF) and the Swiss Nanoscience Institute (SNI). E.M. and R.P. acknowledge funding from the European Research Council (ERC) under the European Union’s Horizon 2020 research and innovation programme (ULTRADISS grant agreement No 834402 and supports as a part of NCCR SPIN, a National Centre of Competence (or Excellence) in Research, funded by the Swiss National Science Foundation (grant number 51NF40-180604). S.-X. L. acknowledges the grant from the SNF (200021\_204053). X.W. and U.A. acknowledge funding by the SNF Professorship (Grant No. PP00P2\_187185/2). Calculations were performed on UBELIX (http://www.id.unibe.ch/hpc), the HPC cluster at the University of Bern. X.L. acknowledges the grants from Natural Science Foundation of Zhejiang Province (LQ22B040003) and National Natural Science Foundation of China (22105172).
\end{acknowledgement}

\begin{suppinfo}
The Supporting Information is available free of charge at !!!.\\

\noindent
Additional STM images and STS data of the $\alpha$ and $\beta$ superlattices; molecular assemblies obtained using DFT and deep learning neural network (DPNN); field-emission resonance tunneling spectra; mechanical dissipation above the pristine Ag(111); simulation of a 1D-periodic array of quantum wells using a periodic Kronig-Penney model; details on the fit procedure of the $\Delta f$(V) spectra considering the quantum capacitance; extraction of the tunneling rates from the dissipation data; determination of the lever arm and additional references (PDF) 
\end{suppinfo}


\end{document}